\documentclass[aps,showpacs,longbibliography,notitlepage,superscriptaddress,twocolumn]{revtex4-1}
\pdfoutput=1
\usepackage[utf8]{inputenc}
\usepackage[english]{babel}
\usepackage[T1]{fontenc}
\usepackage{amsmath}
\usepackage{enumitem}
\usepackage{xcolor}
\colorlet{myPurple}{blue!40!red}
\colorlet{myCyan}{cyan!60!gray}
\colorlet{myRed}{blue!55!gray}
\usepackage{tikz}
\usepackage{pgfplots}
\pgfplotsset{compat=1.14}
\usepackage[colorlinks=true,citecolor=myRed,urlcolor=myRed,linkcolor=myRed]{hyperref}
\usepackage{exscale}
\usepackage{fouriernc}
\usepackage{bbm}
\usepackage{graphicx}
\usepackage{amsmath}
\usepackage{latexsym}
\usepackage{amsfonts}
\usepackage{amssymb}
\usepackage{times}
\usepackage[T1]{fontenc}
\usepackage{amsthm}
\usepackage{enumerate}
\usepackage{bbold}
\usepackage{color}
\usepackage{nicefrac}
\usepackage{changes}
\usepackage{caption}
\usepackage{subcaption}
\usepackage{soul}
\newcommand{\sket}[1]{{\ensuremath{\lvert#1\rangle}}}
\newcommand{\lket}[1]{{\ensuremath{\left\lvert#1\right\rangle}}}
\newcommand{\ket}[1]{\if@display\lket{#1}\else\sket{#1}\fi}

\newcommand{\sbra}[1]{{\ensuremath{\langle#1\rvert}}}
\newcommand{\lbra}[1]{{\ensuremath{\left\langle#1\right\rvert}}}
\newcommand{\bra}[1]{\if@display\lbra{#1}\else\sbra{#1}\fi}

\newcommand{\sbraket}[2]{{\ensuremath{\langle#1\rvert#2\rangle}}}
\newcommand{\lbraket}[2]{{\ensuremath{\left\langle#1\!\left\rvert\vphantom{#1}#2\right.\!\right\rangle}}}
\newcommand{\braket}[2]{\if@display\lbraket{#1}{#2}\else\sbraket{#1}{#2}\fi}

\newcommand{\sketbra}[2]{{\ensuremath{\lvert #1\rangle\!\langle #2\rvert}}}
\newcommand{\lketbra}[2]{{\ensuremath{\left\lvert #1\right\rangle\!\!\left\langle #2\right\rvert}}}
\newcommand{\ketbra}[2]{\if@display\lketbra{#1}{#2}\else\sketbra{#1}{#2}\fi}
\usepackage{tcolorbox}

\newcommand{\proj}[1]{\ketbra{#1}{#1}}

\newcommand{\tr}{\textrm{Tr}}

\usepackage{tikz}
\usepackage{lipsum}
\theoremstyle{plain}

\usepackage{graphicx}
\usepackage{bm}
\usepackage{dsfont}
\usepackage{tikz}
\usepackage[T1]{fontenc}
\usepackage{amsthm}
\usepackage{array}
\usepackage{amssymb}
\usepackage{amsfonts}
\usepackage{cancel}
\usepackage[toc,page]{appendix}
\usepackage{multirow}
\usepackage{color}
\usepackage{calrsfs}
\usetikzlibrary{backgrounds,decorations.pathreplacing,calc}

\definecolor{mygray}{gray}{0.8}
\DeclareMathAlphabet{\mathcal}{OMS}{cmsy}{m}{n}

\begin{document}

\title{Sample-efficient device-independent quantum state verification and certification}

\author{Aleksandra {Go\v{c}anin}}
\affiliation{Faculty of Physics, University of Belgrade, Studentski Trg 12-16, 11000 Belgrade, Serbia}
\author{Ivan \v{S}upi\'{c}}
\affiliation{CNRS, LIP6, Sorbonne Universit\'{e}, 4 place Jussieu, 75005 Paris, France}
\affiliation{D\'{e}partement de Physique Appliqu\'{e}e, Universit\'{e} de Gen\`{e}ve, 1211 Gen\`{e}ve, Switzerland}
\author{Borivoje Daki\'{c}}
\affiliation{University of Vienna, Faculty of Physics, Vienna Center for Quantum Science and Technology, Boltzmanngasse 5, 1090 Vienna, Austria}
\affiliation{Institute for Quantum Optics and Quantum Information (IQOQI),
Austrian Academy of Sciences, Boltzmanngasse 3, 1090 Vienna, Austria}

\date{\today}

\begin{abstract}
Authentication of quantum sources is a crucial task in building reliable and efficient protocols for quantum-information processing. Steady progress vis-\`{a}-vis verification of quantum devices in the scenario with fully characterized measurement devices has been observed in recent years. When it comes to the scenario with uncharacterized measurements, the so-called black-box scenario, practical verification methods are still rather scarce. Development of self-testing methods is an important step forward, but these results so far have been used for reliable verification only by considering the asymptotic behavior of large, identically and independently distributed (IID) samples of a quantum resource. Such strong assumptions deprive the verification procedure of its truly device-independent character. In this paper, we develop a systematic approach to device-independent verification of quantum states free of IID assumptions in the finite copy regime. Remarkably, we show that device-independent verification can be performed with optimal sample efficiency. {Finally, for the case of independent copies, we develop a device-independent protocol for quantum state certification:} a protocol in which a fragment of the resource copies is measured to warrant the rest of the copies to be close to some target state.

%Furthermore, we develop device-independent protocols for quantum state certification {under assumption of independent copies}: a protocol in which a fragment of the resource copies is measured to warrant the rest of the copies to be close to some target state. 

\end{abstract}

\maketitle

\section{Introduction}

Entangled states have been identified as a key resource in applications of quantum technologies, such as quantum communication \cite{repeaters}, computation \cite{ruv},  cryptography \cite{Gisin_2002}, and sensing \cite{Giovannetti_2011}. Harvesting a complete quantum advantage demands an increased precision in manufacturing the relevant components and devices. However, realistic quantum devices are rarely perfect: they suffer from unavoidable errors, noise or decoherence and in some cases, they are not trusted. Therefore, characterization and validation play a central role in real applications of quantum technologies. Verification techniques have a multifaceted complexity. Complexity is measured in terms of a minimal number of queries or instances of resource used (sample complexity), a different number of local measurement settings needed for the procedure (measurement complexity), quantum computational power needed to prepare verification circuit (quantum computational complexity), or classical resources needed to postprocess the verification results (postprocessing complexity). A compact comparison of different techniques in terms of complexity can be found in Ref. \cite{Eisert_2020}.

Complexity directly depends on required confidence and informativeness \cite{Eisert_2020, PRXQuantum.2.010201}: how much of the information we would like to learn about the underlying system and how certain we want to be about the result. The required informativeness depends on the application, and usually, there is a trade-off between informativeness and complexity. Take, for instance, sampling complexity. A less informative technique is, for example, a simple entanglement verification
\cite{guhne2009entanglement, friis2019entanglement} in which (sample) complexity can be reduced to several copies only \cite{Saggio_2019} or even to a logical minimum of a single-copy verification \cite{Dimic1}. On the other hand, fully informative techniques, such as quantum state tomography \cite{Fano, Leonhardt,Hradil_1997,James_2001,Bisio_2009} exhibit exponential sample complexity. Trading informativeness for complexity results in a reduction of resources. Somewhere in between low complexity (e.g., entanglement verification) and exponential complexity (full tomography), one finds resource-efficient tasks such as direct fidelity estimation \cite{Flammia_2011} and shadow tomography tasks \cite{huang2020predicting, morris2019selective, aaronson2019shadow}. Another important example is quantum state verification, which can be achieved with remarkably low sampling complexity \cite{PallisterML_2018,Zhu3,Zhu4}. A specific feature of quantum state verification is that all measurements are local and postprocessing complexity is much lower compared to that of the tomographic methods. The technique received a lot of attention and has been applied to construct verification protocols for various classes of states \cite{Dimic1,Morimae,Takeuchi,Zhu1,Zhu2,Yu_2019,Zhu3,Zhu4,Liu}. It has also been successfully experimentally implemented in Refs. \cite{zhang2019experimental, jiang2020towards}. This task is the central focus of our work.

One drawback of the techniques mentioned so far is that they all assume perfect characterization of all the measurements implemented during the process. This is inconsistent with a device-independent (DI) scenario in which all devices, including the measurement ones, are treated as black boxes~\cite{colbeck,pironio2010random,diqkd,scarani}. Such a paradigm is very useful in the context of protocols involving any potential adversarial activity. A prominent candidate for device-independent state verification is the self-testing method~\cite{STreview,MY}. However, self-testing conclusions are drawn from the knowledge of the asymptotic behavior of an experiment, which in practice invokes the identically and independently distributed (IID) assumption: IID experimental runs. For this reason, self-testing is more of a theoretical prerequisite for practical DI quantum state verification. The literature concerning self-testing usually does not discuss the sample complexity of the procedure. The finite statistics effects in self-testing {in the  non-IID setting} are rigorously treated in Ref.~\cite{JD} on the example of self-testing the maximally entangled pair of qubits. 
{This is closely related to conditional estimators of the Bell inequality violation used in the hypothesis test of local realism without the IID assumption \cite{barrett2002quantum,gill2001accardi,zhang2011asymptotically, bierhorst2015robust, elkouss2016nearly}.}
Other examples for taking into account non-IID effects can be mostly found in the works incorporating self-testing results into protocols for delegated quantum computing such as Refs.~\cite{reichardt2013classical,gheorghiu2017rigidity,coladangelo2019verifier}.

Another downside of verification, in general, is that the measurement process destroys the quantum resource and the conclusion is made about the resource, which is fully consumed.
This hinders the possibility to use the resource for other protocols, without invoking further assumptions. {A} certification protocol, as we define it, represents a way out: a fragment of copies is measured and a conclusion is drawn about the remaining copies. Protocols of this kind have been used for entanglement certification \cite{Arnon_Friedman_2019}, quantum state verification \cite{wu2020efficient}, authentication of teleportation \cite{anu}, quantum contract signing \cite{Yadav_2019}, and single-copy fidelity certification \cite{Zhu3, Zhu4}. 

In this paper, we develop a systematic way to build practical DI quantum state verification protocols starting from self-testing results. Our verification protocol achieves optimal sample efficiency by taking into account the finite statistics effect and is free of the IID assumption. With our methods, self-testing tools find direct application even in cryptographic protocols, where earlier they were deemed inappropriate because of the IID assumption they invoke.
Based on these results we design a DI certification protocol in which a certificate about the unmeasured copies can be guaranteed based on the DI verification procedure performed on the measured copies. This is done by joining together insights from quantum state verification methods with those native to self-testing, and building a general protocol for sample-efficient DI quantum state verification and certification.

\begin{figure*}
\centering
   \includegraphics[width=0.9\textwidth]{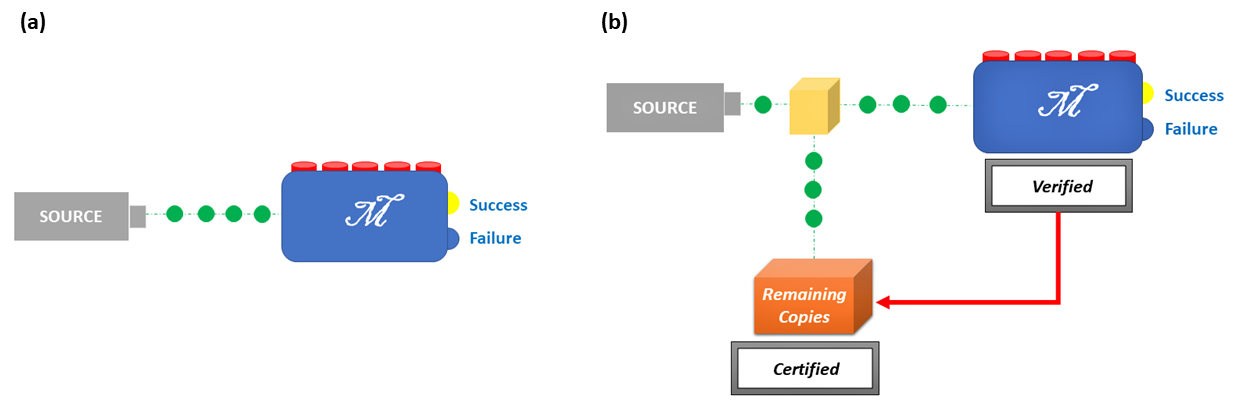} 
       \caption{Comparison of quantum state verification and quantum state certification. The crucial difference is that the former provides a conclusion only after completely consuming the resource, while the latter provides a certificate about the unmeasured resource. For untrusted measurements and the black box scenario we speak of DI quantum state verification and certification.\\ (a) Protocol for quantum state verification. In quantum state verification all available copies are measured and the sample is verified to be (or not) close to the target state. (b) Protocol for quantum state certification. In quantum state certification a part of the sample is measured, and if verified, the rest of the sample is certified to be (or not) close to the target state.}
       \label{fig1}
\end{figure*}

\section {Preliminaries}\label{sec:scene}

In this section, we clarify the terminology and notation and define rigorously the verification and certification tasks, which will be explained in the next sections. Furthermore, we introduce appropriate figures of merit.

\subsection{Verification VS certification}

Typically, the terms verification and certification are used interchangeably in the literature \cite{Eisert_2020}. Here, we distinguish these two terms and associate them to two different tasks. Both tasks refer to the situation in which specification of certain quantum property is confirmed or refuted.

Quantum verification refers to the task in which a certain property of a quantum system is checked, i.e., verified.
For example, a source emits $N$ copies, and the aim is to confirm or refute that the states of the copies are to some degree close to the referent state, by performing appropriate measurements on all of them (Fig. \ref{fig1}(a)). 
Importantly, all the copies produced by the source are consumed during the verification process. 

Quantum certification refers to the scenario in which a subset of the emitted copies is measured, but a certificate of some property of the rest, e.g., proximity to the target state, can still be invoked. 
If the certification is successful, there is a certificate referring to the state of the unmeasured copies (Fig. \ref{fig1}(b)). 

\subsection{Figures of merit}

\textit{Verification.}
In order to quantify the quality of the verification and certification procedure, one needs to choose an appropriate figure of merit. The scenario we consider is the following: a source is producing $N$ independent copies of the quantum system $S = \{\sigma_1,\cdots,\sigma_N\}$, where $\sigma_i$ is the state of the $i$th copy. For now, we abandon the assumption that the copies are identical but still assume that they are independently distributed. The generalizations to the full non-IID case will follow by the end of the next section. In general, our goal is to verify and certify whether the source produces the copies close to some target state $\ket{\psi}$. Formally, the aim is to infer, with some confidence level $1-\delta$, that the average state fidelity of the states from $S$ with  $\ket{\psi}$
\begin{equation}\label{eq:avfid}
F_{\mathrm{av}}(S,\psi) = \frac{1}{N}\sum_{j=1}^N\bra{\psi}\sigma_j\ket{\psi}
\end{equation}
is larger than some value $1-\eta$, with $\eta \in (0,1)$. 

The closeness of the average state fidelity ensures that any (local) measurement on the emitted copies will give statistics
close to the target state statistics.
To see this, let $\Pi$ be an arbitrary measurement operator, $\psi = \ketbra{\psi}{\psi}$ and $ ||\cdot,\cdot||_1$ the trace distance. The following chain of inequalities holds:
\begin{align*}
    \frac{1}{N}\sum_{j=1}^N\tr\left[\Pi(\sigma_j-\psi)\right] &\leq \frac{1}{N}\sum_{j=1}^N ||\psi,\sigma_j||_1\\
    & \leq \frac{1}{N}\sum_{j=1}^N\sqrt{1-F(\psi,\sigma_j)}\\
    &\leq \sqrt{1-F_{\mathrm{av}}(S,\psi)}.
\end{align*}
The first inequality is the property of the trace distance, while the second follows from the relation between trace distance and fidelity \cite{bible}. The last line is obtained by using Jensen's inequality. From these bounds, we see that the average mean is close to the target mean value for arbitrary measurement. Consequently, we know from the Chernoff bound for independent variables that sample mean concentrates around the average mean \cite{chernoff1952measure, hoeffding1994probability}, which in turn implies that local measurement on the sample will give approximately the same statistics as the target state.

However, in a device-independent scenario, one cannot hope to verify fidelity with some particular state. This is due to the fact that the device-independent scenario cannot detect potential local isometries, so a pair of states related by one local isometry can perform equally well in all device-independent tasks. More details about this are given in the next section. The figure of merit has to be fidelity optimized over all local isometries applied to one of the states. In the device-independent literature, this quantity is called \emph{extractability} \cite{bardyn,Jed}. Formally, it is defined as follows. Let $\mathcal{H}_{j}$ be a Hilbert space native to the $j$-th emitted state and let $\mathcal{D}(\mathcal{H}_{j})$ be the set of all density operators on it. We also define the ancillary Hilbert space $\mathcal{H}_{a}$, which is the same as the Hilbert space native to the target state, and the set of density operators on it $\mathcal{D}(\mathcal{H}_{a})$.  Consider a state  the source emits in $j$-th round $\sigma_j \in \mathcal{D}(\mathcal{H}_{j})$, an arbitrary local isometriy $\Phi: \mathcal{D}(\mathcal{H}_{j}) \longrightarrow \mathcal{D}(\mathcal{H}_{j} \otimes  \mathcal{H}_{a})$ and a target state $\psi \in \mathcal{D}(\mathcal{H}_{a})$. Then the extractability of $\psi$ from $\sigma_j$ \footnote{The standard terminology is {`}extractability of the target state from the physical state'. When it is clear from the context what is the target state we will use simply {`}extractability'.} is given by
\begin{equation}
    \label{extract}
    \Xi(\sigma_j,\psi) = \max_{\Phi}F\left(\tr_{j}\left[\Phi(\sigma_j)\right],\psi\right).
\end{equation}
For a sequence of states $S = \{\sigma_1,\cdots,\sigma_N\}$ we define the average extractability of $\psi$ from $S$ as
\begin{equation}
    \label{averagextract}
    \bar{\Xi}(S,\psi) {\stackrel{\textrm{def}}{=}} \frac{1}{N}\sum_{j=1}^N\Xi(\sigma_j,\psi).
\end{equation}
For all practical purposes, the extractability is DI equivalent of fidelity, as in a device-dependent scenario, it simply reduces to fidelity. 
As we commented above, if some state has high fidelity with a given pure target state it means that performing an arbitrary measurement on these two states results in approximately the same statistics. High extractability of the target state from the physical state means that there is an isometry bringing the physical state close to the target state. This isometry has to be taken into account when comparing the measurement statistics, {i.e.}, the statistics obtained by applying an arbitrary measurement to the target state will be approximately the same as the statistics obtained when the image under the inverse isometry of that same measurement is applied to the physical state~\footnote{Note that we can talk about inverse isometry only if the isometry is bijection, hence the statement made here is valid only for bijective isometries. In practice, local isometries used in self-testing are bijections, more precisely they can be seen as local unitaries mapping the physical state in tensor product with a set of ancillas to the target state in a tensor product with some junk state.}.

\textit{Certification.}
In the certification process, the source produces the set of independent states $S = \{\sigma_1,\cdots,\sigma_N\}$. A fraction $N_1 \approx \mu N$, is randomly chosen to be measured, where $\mu$ is the probability for one copy to be measured. Let us denote the set of labels of measured copies with $I$ and the corresponding set of states with $S_v=\{\sigma_i|i\in I\}$. The remaining $N_2 = N - N_1$ copies, we label with $J$, and they are preserved as certificate $S_c = \{\sigma_j|j\in J\}$.
We perform the verification task on $S_v$ and if the test passes we certify (with certain confidence) that $S_c$ has the average fidelity with the target state
\begin{equation}\label{eq:certavfid}
\bar{F}_{\mathrm{cer}}(S_c,\psi) = \frac{1}{N_2}\sum_{j \in J}\bra{\psi}\sigma_j\ket{\psi},
\end{equation}
higher than some value $1-\eta$. To adapt this figure of merit to the device-independent scenario we optimize it over all local isometries and obtain the average extractability of the target state from the remaining copies:
\begin{equation}\label{eq:certavextract}
\bar{\Xi}_{\mathrm{cer}}(S_c,\psi) = \frac{1}{N_2}\sum_{j \in J}\Xi(\sigma_j,\psi).
\end{equation}
So, device-independent certification aims to show that if a test we make passes, the average extractability of the target state from the unmeasured copies must be higher than some value with a given confidence level.

\textit{Non-IID case.}
Up to now, we assumed that the copies produced by the source are mutually independent, i.e., there are no correlations among them. To indicate how correlations can hinder the verification process, let us consider the source that over $N$ rounds produces the full state $0.5\ketbra{\psi}{\psi}^{\otimes N} + 0.5\chi^{\otimes N}$, where $\ket{\psi}$ is the target state and $\chi$ is some separable state with very low fidelity with $\psi$.
Such a source is not reliable as it has a $50\%$ chance to produce a "bad" state in all rounds. In an actual experiment the source would emit either the sequence $\{\ket{\psi},\cdots,\ket{\psi}\}$ or the sequence $\{\chi,\cdots,\chi\}$. The first one would pass all verification tests, and based on such a test one might wrongly conclude that the source is good.
{To circumvent this behavior of non-IID sources, one way out is to verify not the correct functionality of the source, but the sequence generated in the specific experiment. To formalize the former, one introduces conditional states that actually correctly capture the above example. Of course, one may potentially formalize non-IIDness in a different way, unfortunately there is not much literature on this topic and this is left for future considerations.}

%{In order to describe experimental results correctly and capture the nature of the quantum state describing all $N$ copies, one must introduce a quantity that encloses both sequential nature of the experiment and potential mutual dependence of the copies.} 
%To remedy this, 
In this work, we adhere to the idea of conditional fidelity, introduced in Ref. ~\cite{JD}. With the aid of the techniques introduced there we can consider the verification process also in the case when the full state produced by the source $\sigma^N$ is correlated, or even entangled. In that case, in every round we consider the conditional state
\begin{equation}\label{eq:condst}
  {  \tilde{\sigma}_{j} = \frac{1}{p_{1,\cdots,j-1}}\tr_{1,\cdots,j-1,j+1,\cdots,N}[(\otimes_{k=1}^{j-1} M_{\mathbf{o}_k|\mathbf{i}_k})\sigma^N],}
\end{equation}
where $M_{o_k|i_k}$ is the measurement performed on the $k$-th copy, while $p_{1,\cdots,j-1} = \tr[(\otimes_{k=1}^{j-1} M_{\mathbf{o}_k|\mathbf{i}_k})\sigma^N]$. Thus, the sequence of states $\tilde{S} = \{\tilde{\sigma}_1,\cdots,\tilde{\sigma}_N\}$ consists of mutually independent copies, and it represents the sequence of states actually measured in this particular experiment of $N$ runs. {The conditional states trivially encompass  the IID and independent case scenario (per definition). Clearly, they also correctly capture the example provided above, i.e. the test fails if the ``bad'' state $\chi^{\otimes N}$ is produced, while the test passes if the sequence of target states is generated.} The figure of merit in this case is the average conditional fidelity
\begin{equation}\label{eq:noniidavfid}
\tilde{F}_{\mathrm{cond}}(S,\psi) = \frac{1}{N}\sum_{j=1}^N\bra{\psi}\tilde{\sigma}_j\ket{\psi}
\end{equation}
and its device-independent counterpart, the average conditional extractability
\begin{equation}\label{eq:noniidavextract}
\bar{\Xi}_{\mathrm{cond}}(S,\psi) = \frac{1}{N}\sum_{j=1}^N\Xi(\tilde{\sigma}_j,\psi).
\end{equation}

Hence, the non-IID verification process at the end returns the average extractability of the target state $\ket{\psi}$ from the sequence of the conditional states $\tilde{S}$.

In the following sections we design the first {general} protocol for DI quantum state verification, both in the scenario where different experimental rounds are independent but not identical, and in the full non-IID scenario. {As far as we know the only rigorous treatment of non-IID and DI quantum state verification was presented in Ref. \cite{JD} for the particular case of maximally entangled pair of qubits. Our method is general, and can be used to verify any quantum state that can be self-tested.} Concerning the DI quantum state certification we propose the first protocol in the scenario with independent experimental rounds, and comment on the conceptual difficulties when it comes to designing the full non-IID protocol.

\section{A framework for device-independent quantum state verification}

In this section, we build the protocol for non-IID quantum state verification. We start by discussing two main ingredients for the protocol, which are quantum state verification, Sec.~\ref{sec:qsv}, and self-testing Sec.~\ref{sec:st}. Then we show how they can be used to construct a DI protocol for quantum state verification and present a few case studies.

\subsection{Quantum state verification}\label{sec:qsv}

\emph{Independent copies.} In this section, we recall the framework for quantum state verification \cite{pallister_2018, PallisterML_2018,Zhu3,Zhu4}. The aim is to verify that an uncharacterized source is producing the target state $\sigma = \ketbra{\psi}{\psi}$ by using only local measurements.
Here, the source is assumed to produce a sequence of mutually independent copies $S = \{\sigma_1,\cdots,\sigma_N\}$.

The verification procedure consists of performing fully characterized measurements on the copies produced by the source. The measurement strategy $\Omega$, thus consists of $l$ different binary local measurements $\{M_{o|i}\}$, where measurement settings are labeled with $i \in \{1,\cdots, l\}$ and outputs with $o \in \{0,1\}$.  In the $j$th round, a label $i$ is randomly sampled and the corresponding measurement is applied to the state $\sigma_j$. We say that the state $\sigma_j$ has passed the round if it returned the output $o = 1$. Otherwise, we say it failed. The first time a round fails, the process is aborted. The measurements are chosen to be such that $\ketbra{\psi}{\psi}$ is the only state with $\bra{\psi}M_{1|i}\ket{\psi} = 1$ for all $i = 1,\cdots,l$. In other words, only the target state has $100\%$ chance to successfully pass all the rounds. If all $N$ rounds are passed we verify with certain confidence level $1-\delta$ that the average fidelity $F_{\mathrm{av}}(S,\psi)$, defined in Eq. \eqref{eq:avfid}, must be higher than some value $1-\eta$:
\begin{equation}
   F_{\mathrm{av}}(S,\psi) > 1 - \eta.
\end{equation}

The analysis is based on finding the second-largest eigenvalue of the strategy operator $\hat{\Omega} = \sum_ip_iM_{1|i}$, where $p$ usually corresponds to a uniform distribution. The largest eigenvalue of $\hat{\Omega}$ is $1$ and the corresponding eigenvector is $\ketbra{\psi}{\psi}$. Of all states $\rho$ with $\bra{\psi}\rho\ket{\psi} = 1 - \eta$, the highest average probability to pass a randomly chosen round has the state $(1-\eta)\ketbra{\psi}{\psi} + \eta\ketbra{\psi'}{\psi'}$, where $\ket{\psi'}$ is the eigenvector of $\hat{\Omega}$ associated to the second-largest eigenvalue $1 - \nu(\Omega)$ {[here $\nu(\Omega)$ is defined as the gap between the largest and the second largest
eigenvalue of $\hat{\Omega}$]} , resulting in the probability of $1-\eta\nu(\Omega)$ to pass the round in the best case. Thus, for $N$ rounds, the best strategy is achieved when all states have exactly the same fidelity with $\ket{\psi}$ \cite{Zhu3} resulting in overall success probability $[1-\eta \nu(\Omega)]^N$. Our aim is to confirm that the average fidelity of the states from $S$ with $\ket{\psi}$ is larger than $1-\eta$. Hence, to verify the target state with the confidence level $1-\delta$,

\begin{equation} \label{localstrategy}
N\geq {\frac{\ln{\delta}}{\ln[1-\eta \nu(\Omega)]}} \approx \frac{1}{\eta\nu(\Omega)}\ln\delta^{-1},
\end{equation}
which scales as $N=O[1/\nu (\Omega)\eta]$.

Note that the efficiency of the procedure depends on the spectral gap of the strategy operator $\hat{\Omega}$. In theory, the largest gap is equal to $1$, in which case the strategy consists of a single measurement ${\ketbra{\psi}{\psi}, 1- \ketbra{\psi}{\psi}}$, {and for which one gets optimal scaling of $N=O(1/\eta)$.}
However, such a measurement is entangled whenever $\ket{\psi}$ is, and for the sake of experimental simplicity, one of the requirements of the protocol is that the strategy involves only local measurements. Surprisingly, Eq. \eqref{localstrategy} shows that in certain cases we can achieve the same optimal scaling for local strategies (up to a constant factor). One can generalize the procedure to nonperfect strategies characterized by the strategy operator $\hat{\Omega}$, such that $\tr[\hat{\Omega}\psi] < 1$ and obtain quadratically worse scaling \cite{pallister_2018}.
In that case, the procedure does not halt the first time a round is failed. After $N$ rounds the lower bound of the average fidelity is estimated based on the frequency of the successful rounds. This is particularly important for creating DI verification protocols, as for some states there are no self-tests that can be phrased as tasks passed by the target state with maximum probability. In this case, we get quadratically worst scaling of the number of copies ${O}(1/\eta^2)$, which we call suboptimal scaling, but still efficient in terms of resources.

\emph{Non-IID case.} Until now, adapting the quantum verification procedure to a non-IID scenario has not received a lot of attention. 
So far, recent work \cite{Zhu3} presented the method for building verification protocols adapted to such a scenario, starting from an already existing one for the independent copies \cite{Zhu4}. In their work, authors construct a non-IID protocol in which measuring $N-1$ copies allows the fidelity certificate for the remaining unmeasured one to be proved.
This procedure is discussed in more details in Sec. \ref{DIQSC}. In our work, we implement a DI scenario and go beyond one-copy certification with the goal to certify an asymptotically large number of copies $O(N)$.

\subsection{Self-testing}\label{sec:st}

Self-testing is the only known device-independent verification procedure \cite{MY,STreview}. It aims to verify that a source is producing a certain target state ${\ket{\psi}}$. The state actually produced by the source is called the physical state and the applied measurements are termed physical measurements. The key aspect of self-testing is device-independence: local measurements are not characterized, nor trusted. All devices are treated as black boxes. Different parties, sharing a multipartite state ${\sigma}$ are spatially separated and noncommunicating. Each party queries their box with classical input $i$, denoting the measurement choice, and the boxes return classical output $o$. Usually, the sources are assumed to be IID. Therefore, by repeating the measurement process for many rounds the parties can estimate the probabilities of obtaining the set of outputs $o_1,o_2,\cdots, o_n$ when the set of inputs is $i_1,i_2,\cdots,i_n$:
\begin{equation*}
    \{p(o_1,o_2,\cdots, o_n|i_1,i_2,\cdots, i_n)\}
\end{equation*}
The set of probabilities is usually termed, simply, correlations. Assuming the correctness of quantum mechanics, the correlations are related to the physical state and measurements through the Born rule:
\begin{equation}\label{eq:born}
    p(o_1,o_2,\cdots, o_n|i_1,i_2,\cdots, i_n) = \tr\left[\bigotimes_{j=1}^n M_{o_j|i_j}\sigma\right]. 
\end{equation}
As all information is drawn only from the correlation probabilities it is impossible to verify the exact form of the physical state. Indeed, certain transformations such as simultaneous local rotations of the state and measurements, or embedding in Hilbert spaces of larger dimension, leave the observed probabilities unchanged. These transformations are captured by the notion of local isometries, and the best one can hope for is to verify that a local isometry maps the physical state to the target one. If such local isometry exists one says that the physical and the target state are equivalent. Rigorously written, the states $\sigma \in \mathcal{D}(\mathcal{H}')$ and ${\psi} \in \mathcal{D}(\mathcal{H})$ are equivalent if there exists a local isometry $\Phi:\mathcal{D}(\mathcal{H}') \longrightarrow \mathcal{D}(\mathcal{H}'\otimes\mathcal{H})$ such that
\begin{equation}
    \Phi(\sigma) = \xi\otimes{\psi}.
\end{equation}
In some cases, if the observed correlations are the same as the correlations obtained from the target state, one can infer the equivalence between the physical and the target state. 
In general, the self-testing statement does not need to specify the form of the isometry, but in most cases, the proof is based on the specific isometry.

One of the bigger challenges on the way towards constructing a self-testing protocol is to find candidate correlations. Natural candidates are the correlations reaching the quantum bound of some of Bell's inequalities:
\begin{equation}
    b = \sum_{\substack{o_1,o_2\cdots o_n\\i_1,i_2,\cdots i_n}}\beta_{o_1,o_1,\cdots,o_n}^{i_1,i_2,\cdots,i_n}p(o_1,o_2,\cdots, o_n|i_1,i_2,\cdots, i_n).
\end{equation}
The local bound $b_L$ is the limit achievable by the correlations compatible with local hidden variable models. The quantum bound $b_Q$ is the maximal violation achievable by the correlations compatible with quantum theory, i.e., those for which there exist measurement operators $\{M_{o_j|i_j}\}_{j=1}^n$ and density matrix $\sigma$, satisfying Eq. \eqref{eq:born}. In many cases, the quantum bound can be achieved only by using a specific state named the target state (up to local isometries), which is the basis of self-testing.

The simplest and probably the most widely used in self-testing are the correlations that maximally violate the Clauser-Horn-Shimony-Holt (CHSH) inequality \cite{chsh}. It has been proven that two parties can maximally violate the CHSH inequality only if they apply anticommuting measurements on a maximally entangled state of two qubits. Thus, the CHSH inequality can be used to self-test the maximally entangled pair of qubits.

Let us now concentrate on quantum bounds that are also algebraic bounds, i.e. bounds constrained only by the positivity of observed probabilities.
Notably, for a fixed set of inputs $\mathbf{i} = (i_1,i_2,\cdots,i_n)$, we define $\beta^\mathbf{i} =\max_{o_1,o_2,\cdots,o_n} \{\beta_{o_1,o_2,\cdots,o_n}^{i_1,i_2,\cdots,i_n}\}$. If the quantum bound of the inequality is equal to the algebraic bound, the probabilities  $p(o_1,o_2,\cdots, o_n|i_1,i_2,\cdots, i_n)$, associated with coefficients $\beta_{o_1,o_2,\cdots,o_n}^{i_1,i_2,\cdots,i_n} = \beta^\mathbf{i}$  have to sum to $1$. Probabilities associated to the coefficients $\beta_{o_1,o_2,\cdots,o_n}^{i_1,i_2,\cdots,i_n} < \beta^\mathbf{i}$ are equal to $0$. This holds for all sets of inputs. An example of a Bell inequality whose quantum bound is equal to the algebraic one and self-tests the underlying state is the Mermin inequality \cite{Mermin}. All states maximally violating it have to be equivalent up to local isometries to the Greenberger-Horn-Zeilinger (GHZ) state $\ket{\psi_{GHZ}} = (\ket{000}+\ket{111})/\sqrt{2}$.

A very important aspect of a self-testing procedure is its robustness. Let us consider a self-test based on the maximal violation of a certain Bell inequality. The maximal violation is very difficult to reach due to the unavoidable presence of noise and imperfections. It is desirable that one can establish a lower bound on the extractability of the target state from the physical state in the case when maximal violation is not exactly reached.

A self-test is robust if observing violation $b < b_Q$ allows putting a lower bound on the extractability of the target state from the physical one.
\textit{The extractability $\Xi(\sigma,\ket{\psi})$ of the target state $\psi$ from any state $\sigma$ achieving the Bell violation $b_Q-\tilde{\varepsilon}$ is at least $1-f(\tilde{\varepsilon})$, where $f$ is a function, which depends on the characteristics of the given self-test.}  As shown in Ref. \cite{Jed}, a {lower bound to the} extractability can always be represented as a linear function of the Bell violation, i.e., without loss of generality we take $f(\tilde{\varepsilon})= \tilde{c}\tilde{\varepsilon}$ that $\tilde{c}$ is a constant. Furthermore, if the isometry $\Phi$ used in the self-test is optimal, we say that the self-test is {\emph{tight}} in terms of robustness. Such is the self-test of the GHZ state based on the violation of the Mermin inequality \cite{Jed}.

\subsection{Device-independent quantum state verification}\label{sec:diqsv}

The previous two sections gave us all the ingredients for constructing a protocol for DI quantum state verification. For clarity, we start with the case where the target state for the self-test achieves the algebraic bound of the corresponding Bell inequality, and we discuss the general case afterwards. The verification can be performed in two different setups, i.e., in the case of independently distributed copies and the case of non-IID source. Both scenarios aim to construct a DI verification procedure characterized by the optimal sample efficiency.

\textit{Independently distributed copies.} As the procedure is DI the aim is to verify whether the \textit{average} extractability $\bar{\Xi}(S,\psi)$ of the target state $\ket{\psi}$ from the set of states $S=\{\sigma_1,\sigma_2,...,\sigma_N\}$
\begin{equation}
    \bar{\Xi}(S,\psi) = \frac{1}{N}\sum_{j=1}^N \Xi(\sigma_j,\ket{\psi})
\end{equation}
is above some value or not. In the language of hypothesis testing, our hypothesis is the following: the average extractability $\bar{\Xi}(S,\psi)$ of the target state from $S$ is higher than a given value, and we aim to test it. 

The first step is to view the Bell inequality as a procedure in which only the states equivalent to the target state can pass all rounds, i.e., as a nonlocal game. For the criteria and procedure to convert a Bell inequality into the nonlocal game see Ref. \cite{Silman}. For the set of inputs $\mathbf{i} = (i_1,i_2,\cdots,i_n)$ let us denote the outputs corresponding to $\beta^{\mathbf{i}} = \max_{o_1,o_2,\cdots,o_n}\{\beta_{o_1,o_2,\cdots,o_n}^{i_1,i_2,\cdots,i_n}\}$ as the correct outputs, while all the others are denoted as wrong outputs. As only a state equivalent to the target state is able to maximally violate the Bell inequality, such {a} state alone can return the correct outputs for every set of inputs. This already provides us with contours for a DI verification procedure. {The} first time the boxes return a set of wrong outputs the procedure can be halted, since either the underlying state is not equivalent to the target state or the measurements performed are not those allowing construction of a successful verification procedure.

For the second step, we need to estimate the number of copies necessary to exceed a certain bound on the average extractability. For that purpose let us look into the robustness bounds. Conversely, the robustness statement (see the previous section) says that the state $\sigma$ for which there exists isometry $\Phi$, such that $F[\Phi(\sigma),\ket{\psi}]= 1-\eta$, and thus $\Xi(\sigma, \psi)\leq 1-\eta$, at best can achieve the Bell violation  $b_Q - \eta/\tilde{c}$, where $\tilde{c}$ is a constant characterizing linear dependence between extractability and the Bell violation. 
Furthermore, we translate the self-test procedure into a binary task, the so-called nonlocal game, in which local input settings correspond to binary questions
queried at random. The performance is characterized by the percentage of the correct answers, i.e., the winning probability. Thus, a Bell violation $b_Q - \eta/\tilde{c}$ directly translates to probability of success  $1- \eta/b_Q\tilde{c} = 1-\eta c$, where we set $c=1/b_{Q}\tilde{c}$ that we use frequently throughout the text.

Note that, in the case of a nonalgebraic bound, i.e., a Bell inequality with a gap between quantum and algebraic bound, {the} maximal probability of success is $p_{\mathrm{QM}}<1$ ($p_{\mathrm{QM}}$ denotes quantum bound), but relation remains linear. 
This establishes a direct relation between extractability and probability of success in a self-test. Once this is known it is easy to estimate the number of copies necessary to pass the verification test.

After a more general description, it is instructive to look at a specific case. The simplest example is the tripartite qubit GHZ state and self-testing based on Mermin's inequality~\cite{Mermin}
\begin{eqnarray}
&\sum&_{o_1,o_2,o_3}(-1)^{o_1+o_2+o_3}\big[p(o_1,o_2,o_3|0,0,1) + p(o_1,o_2,o_3|0,1,0) + \nonumber \\ &+& p(o_1,o_2,o_3|1,0,0)- p(o_1,o_2,o_3|1,1,1)\big] \leq 2.
\end{eqnarray}
Any violation of the Mermin inequality allows a nontrivial conclusion to be made about the extractability of the GHZ state from the underlying state. In particular, a state $\sigma_j$ achieving violation $4-\tilde{\varepsilon}_j$ has extractability with GHZ higher than
$1 - \tilde{\varepsilon}_j/c$, where { $c=2-\sqrt{2}$}. This bound is proven to be tight \cite{Jed}.
Therefore, we translate the self-test procedure into the nonlocal game and characterize it using probability of success. For example, if in a round we ask one of the global questions [set of inputs $(0,0,1)$, $(0,1,0)$, $(1,0,0)$, or $(1,1,1)$] and get the correct outputs, the achieved score in the round is $p_j = 1$, otherwise $p_j = 0$. The final score is $P = \sum_{j=1}^Np_j/N$.
Since the self-test is tight we can conclude that if extractability of the GHZ state from the state $\sigma_j$ is not greater than $1 - \eta_j$, the highest violation of the Mermin inequality the state $\sigma_j$ can achieve is {$4 - \eta_j/\tilde{c}$}. It further means that the state has probability of success {$p_{\eta_j} = 1-\eta_j/4\tilde{c}= 1-c\eta_j$.}

{The probability that the sequence of states $S$ passes all rounds $\Pi_{j=1}^Np_{\eta_j} = \Pi_{j=1}^N(1-c\eta_j)$ is bounded}
{
\begin{equation}\label{ineq}
\Pi_{j=1}^N(1-c\eta_j) \leq \Pi_{j=1}^N(1-c\eta) = (1-c\eta)^N,
\end{equation}
}
{where $1 - \eta = 1 - \sum_{j=1}^N\eta_j/N$ is the average extractability $\bar{\Xi}(S, \sigma)$. Relation \eqref{ineq} is a consequence between the  inequality of arithmetic and geometric means:}

\begin{align*}
    \frac{1}{N}\sum_{j=1}^N(1-c\eta_j) &= 1 - c\eta\\ &= \sqrt[\leftroot{-3}\uproot{3}N]{(1-c\eta)\times\cdots\times(1-c\eta)} \\ &\geq \sqrt[\leftroot{-3}\uproot{3}N]{(1-c\eta_1)\times\cdots\times(1-c\eta_N)}.
\end{align*}

The bound is saturated when $\eta_j = \eta$ for all $j = 1,\cdots,N$. For the fixed average extractability $1-\eta$, the optimal performance is achieved when the GHZ state has the same extractability $1-\eta$ from all the states belonging to the sequence $S$. In this case, a probability that a sequence of states, characterized by the average extractability of the target state lower than $1-\eta$, gives a correct answer in all $N$ rounds is

\begin{align}
p\left[o_1=...=o_N=1|\bar{\Xi}(S,\psi)\leq 1-\eta\right]\leq (1-c\eta)^N
\end{align}

In other words, the confidence level $1-\delta$ that the average extractability $\bar{\Xi}(S,\psi)$ is larger than $1-\eta$ has a lower bound $1-(1-c\eta)^N$. From here we can estimate the number of copies:  
\begin{equation} \label{confDI}
    N \geq {\frac{\ln\delta}{\ln(1-c\eta)}} \approx \frac{1}{c\eta}\ln\delta^{-1}.
\end{equation}
Importantly, this is asymptotically \textit{the best sample efficiency one can achieve} (up to a constant factor).

Here we discuss the case where all the copies passed the test successfully, which is impossible in real experiments.
For that reason, we now adapt the procedure to a more realistic scenario in which some number of failed rounds is allowed before halting the protocol. Furthermore, in what follows we do not impose that the target state has to be self-testable through a Bell inequality whose quantum bound is equal to the algebraic one. The general protocol for DI quantum state verification can be summarized as follows.
\vskip 0.2cm
\textbf{Protocol 1:} {Device-independent quantum state verification.}
\vskip 0.1cm

 $\mathbf{1)}$ Fix the lower bound on  average extractability that we want to certify $1 - \eta$. This implies the lower bound on the average  success probability of the whole sample $p_{\mathrm{QM}} - \varepsilon_2$, i.e., $p_{\mathrm{QM}}-c\eta$. \vspace{2mm}

 $\mathbf{2)}$ Fix the allowed tolerance from the optimal success rate $\varepsilon_1 < \varepsilon_2$, i.e., $P\geq p_1= p_{\mathrm{QM}}- \varepsilon_1$ and the corresponding verification confidence $1-\delta$.\vspace{2mm}

 $\mathbf{3)}$ {Fix the number of samples $N$ according to the lower bound given in Eq. \eqref{numberofcopiesverification}.}\vspace{2mm}

 $\mathbf{4)}$  Run the protocol: measure all the available copies according to a procedure corresponding to a self-test for the corresponding target state. \vspace{2mm}

 $\mathbf{5)}$   If the success rate $P\geq p_1$, the protocol is successful and average extractability of the measured sequence of states is $\bar{\Xi} \geq 1-\eta$ with confidence level $1-\delta$. Otherwise, the protocol is inconclusive. 
\vskip 0.1cm

We start the explanation of the protocol with the observation that a self-testing robustness bound provides the relation between the success probability and extractability. A state $\sigma_j$ providing extractability $1-\eta_j$ to the target state has maximal achievable success probability $p_{\eta_j} =p_{\mathrm{QM}} -c\eta_j$, where $p_{\mathrm{QM}}$ is the success probability of the target state. 
A given sequence of independent copies $S=\{ \sigma_1,..., \sigma_N\}$ can be characterized with the average extractability $\bar{\Xi}(S,\psi)$, which corresponds to the upper bound on average success probability $\bar{p} {\stackrel{\textrm{def}}{=}} \frac{1}{N}\sum_{j=1}^Np_{\eta_j}$. We say that the verification test is successful if the success rate $P=\frac{1}{N}\sum_{j=1}^N o_j$ passes certain fixed threshold $p_1 = p_{\mathrm{QM}}-\varepsilon_1$. 
The aim is to show that the probability for a sequence of states characterized with average {extractability} $\bar{\Xi}\leq 1-\eta$ to pass the test decreases exponentially with the number of copies. In order to do so, we prove that this holds under stronger assumption, i.e., if success probability $\bar{p}\leq p_{\mathrm{QM}}-\varepsilon_2$ ($\varepsilon_2=c\eta$). The relation between two is as follows. 
{If we denote the set of states satisfying $\bar{\Xi}(S,\psi)\leq 1-\eta$ with $\Sigma_{\eta}$, and the set of states satisfying $\bar{p}\leq p_{\mathrm{QM}}-c \eta$ with $\Pi_{\eta}$, then on the level of sets we have $\Sigma_{\eta}\subseteq \Pi_{\eta}$. Our target quantity is the probability 
\begin{equation}\label{eq:19}
p\left[P\geq p_{\mathrm{QM}}-\varepsilon_1|\bar{\Xi}\leq 1-\eta\right]\stackrel{\textrm{def}}{=} p\left[P\geq p_1|S\in \Sigma_\eta\right].
\end{equation} 
Similarly, we define 
\begin{equation}\label{eq:20}
p\left[P\geq p_1|\bar{p}\leq p_{\mathrm{QM}}-\varepsilon_2\right]\stackrel{\textrm{def}}{=}p\left[P\geq p_1|S\in \Pi_\eta\right].
\end{equation}
Following the set relations one directly reads
\begin{eqnarray}\label{eq:21}
\max_{S \in \Sigma_\eta}\;&p&\left[P\geq p_1|S\in \Sigma_\eta\right]\leq\\
\nonumber  \leq \max_{S\in \Pi_\eta}\;&p&\left[P\geq p_1|S\in \Pi_\eta\right]{\stackrel{\textrm{def}}{=}} p_{\textrm{max}}\left[P\geq p_1|\bar{p}\leq p_{\mathrm{QM}}-\varepsilon_2\right]
\end{eqnarray}
Further on, we derive the upper bound on $p_{\textrm{max}}$ related to the probability of success, which will then automatically imply, by Eq. \eqref{eq:21}, the desired bound on extractability.} This probability can be expressed with the help of the Chernoff-like bound (for details, see Appendix \ref{proof}):

\begin{equation} \label{generalverification}
    p_{\textrm{max}}\left[P\geq p_{\mathrm{QM}}-\varepsilon_1|\bar{p}\leq p_{\mathrm{QM}}-\varepsilon_2\right]\leq \mathrm{e}^{-D(p_1||p_2)N},
\end{equation}
where {$p_2=p_{\mathrm{QM}}-\varepsilon_2$} and $D(a||b) = a\log(a/b) + (1-a)\log((1-a)/(1-b))$ is the Kullback-Leibler (KL) divergence. From here we can estimate the number of copies needed to verify the lower bound of the extractability
\begin{equation} \label{numberofcopiesverification}
N\geq\frac{\ln{\delta^{-1}}}{D(p_{\mathrm{QM}}-\varepsilon_1||p_{\mathrm{QM}}-\varepsilon_2)}
\end{equation}
where $1-\delta$ is the required confidence level.
Note that, if the test is passed, the condition $\bar{p}>p_{\mathrm{QM}}-\varepsilon_2$ directly translates to the extractability bound, i.e., extractability must be greater than $1-\epsilon_2/c$.

Now, we see how KL divergence gives two completely different scalings in two regimes.
Let us consider the case when the quantum bound is equal to the algebraic bound $p_{\mathrm{QM}}=1$ and make a Taylor expansion of Eq. \eqref{numberofcopiesverification}. 
In this case, we obtain that the number of copies $N$ scales like $O(\ln{\delta^{-1}}/c\eta)$, which is the optimal scaling. 
On the other hand, if we work with nonalgebraic bound $p_{\mathrm{QM}}<1$, Taylor expansion of Eq. \eqref{numberofcopiesverification} gives $N=O(\ln{\delta^{-1}}/c^2\eta^2)$. All details of the expansions can be found in Appendix \ref{proof}.

Finally, with this we can formulate our first result.
\vskip 0.2cm
\textbf{Result 1:} DI quantum state verification.
\vskip 0.1cm
\emph{The entangled state $\psi$ can be device-independently
verified if there exists a robust self-test for it, based on the set of correlations, which can be phrased as a nonlocal game. The verification test achieves optimal sample efficiency of $O(\ln{\delta^{-1}}/c\eta)$ if the state saturates the algebraic bound of a corresponding Bell inequality. For nonalgebraic bound self-tests, the state is verified with suboptimal sample efficiency of $O(\ln{\delta^{-1}}/c^2\eta^2)$.}
\vskip 0.1cm

 {In comparison with the device-dependent case, see Eq. \eqref{localstrategy} and comments below, obtained results for the device-independent case differ only by a constant factor, both for optimal and suboptimal scaling.}
\begin{figure*}
    \centering
    \includegraphics[width=0.95 \textwidth]{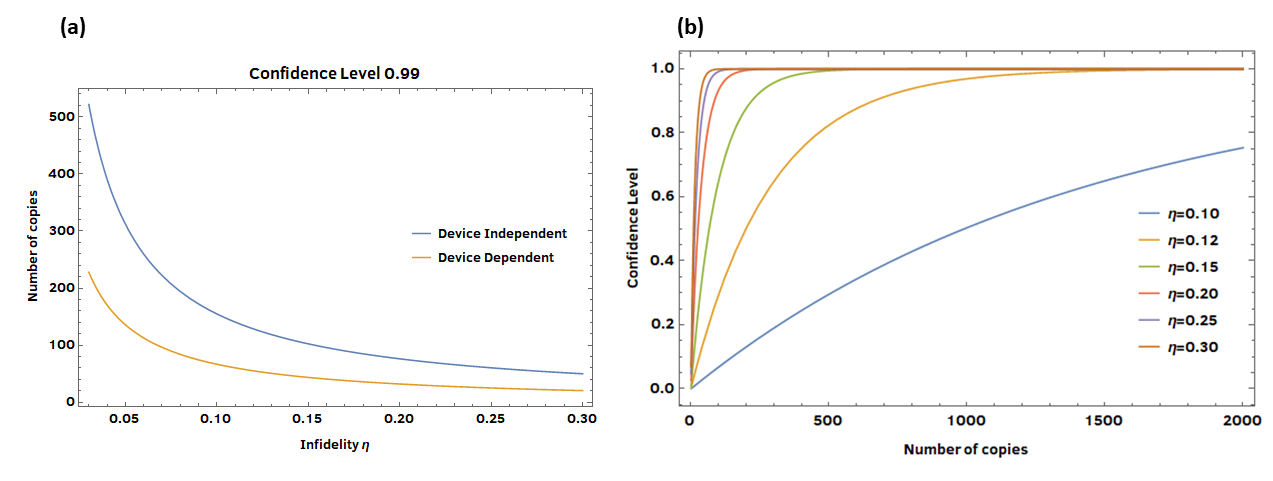}
    \caption{Examples of DI quantum state verification. (a) The verification of the tripartite GHZ state. We compare the number of state copies $N$ needed for quantum state verification in the device-independent, Eq. \eqref{confDI}, versus device-dependent scenario for a fixed confidence level \cite{li2020optimal}. The scaling of the device-dependent version of verification is better only by a constant factor, which in the particular example has the value $2(2+\sqrt{2})/3$. This is the type of test where we assume that all rounds are successfully passed {(i.e., $p_1=1$ in our protocol 1).}\\ (b) Verification of tripartite GHZ state in a realistic scenario. For different values of $\eta$, where $1-\eta$ is the average extractability, we compare the growth of the confidence level as a function of the number of copies $N$ of the prepared quantum state under the assumption that on average $95 \%$ of experimental runs are successful (i.e., overall success rate $P$ is greater than $p_1=0.95$).}  
    \label{fig2}
\end{figure*}

In Fig. \ref{fig2}(a) we compare the sample efficiency of device-dependent and device-independent quantum state verification of the tripartite GHZ state. The DI protocol is based on the performance in Mermin's GHZ game. As a device-dependent verification protocol we use the optimal protocol, proposed in Ref. \cite{li2020optimal}. The scaling of the device-dependent version of quantum state verification is better only by a constant factor. This scaling is obtained for the protocol not tolerating any failed rounds, as in the standard quantum state verification.  
The theoretical plot for stabilization of the confidence level for different average fidelities $1-\eta$ in a more realistic scenario is given in Fig. \ref{fig2}(b). In particular, we take that $0.95$ of copies successfully passed experimental runs, therefore the maximal estimated value for extractability is approximately $0.90$ (obtained using linear dependence between fidelity and the average probability of success). 

\textit{Non-IID source.} The previous section accounts for independent copies the source produces. Here we adopt the strategy used in the context of experimental self-testing protocol~\cite{JD} to adapt {our} verification protocol to {the} non-IID case. As indicated in Sec. \ref{sec:scene} the idea is to verify, not the correct functioning of the source, but the average extractability of the target state from the states produced by the source in a given sequence. Let us return to the example from Sec. \ref{sec:scene}: a source "flips a coin" and depending on the outcome decides whether to send a sequence of target states $\ket{\psi}$, or a sequence of useless separable states $\chi$, such that $F(\chi,\psi) = 0.5$. The full state the source produces is  $\rho = (\proj{\psi}^{\otimes n} + \chi^{\otimes n})/2$. Obviously, the source is not good, as the extractability cannot pass $0.75$. However, with probability $50\%$ the source sends the sequence of target states, which can pass the test in all rounds. If this happens, we can indeed verify that the actual produced sequence is good. 
And, for practical purposes, it is a perfectly plausible task: verification of the sequence of the states one had access to, rather than the verification of the source itself.

To formalize this let us introduce the conditioning. Conditional extractability of the target state from the state available in round $j$ is defined in the following way:
{\begin{multline}
    \Xi_{j|\mathrm{past}}(\tilde{\sigma}_j,\ket{\psi}) {\stackrel{\textrm{def}}{=}} \\  \frac{1}{p_{1,\cdots,j-1}}\left\{\Xi\left[\tr_{1,\cdots,j-1}\left((\otimes_{k=1}^{j-1} M_{\mathbf{o}_k|\mathbf{i}_k})\sigma^{[j]}\large\right),\ket{\psi}\right]\right\}
\end{multline}}
where $\sigma^{[j]}$ is the full state over $j$ rounds, $M_{\mathbf{o}_k|\mathbf{i}_k}$ is the measurement operator used in the $k$-th round, $p_{1,\cdots,j-1} = \tr[(\otimes_{k=1}^{j-1} M_{\mathbf{o}_k|\mathbf{i}_k})\rho]$. Average \textit{conditional} extractability for the sequence of states $S$ is defined as
\begin{equation}
    \bar{\Xi}_{\mathrm{cond}}(S,\psi) {\stackrel{\textrm{def}}{=}} \frac{1}{N}\sum_{j=1}^N\Xi_{j|{\mathrm{past}}}({\tilde{\sigma}_j},\ket{\psi}).
\end{equation}

If the states used in different rounds are correlated, as in the example given above, the score in round $i$ can be viewed as conditional, as it might depend on all the previous scores. Of course, the rounds are performed sequentially in order to get adequate conditional probabilities. Such a conditional score in the round $j$ is denoted with $p_{j|\mathrm{past}}$, and importantly all conditional scores are mutually independent by construction: $p_{j+1,j|\mathrm{past}} = p_{j+1|\mathrm{past}}p_{j|\mathrm{past}}$. The final score is average of all the conditional scores $\bar{p} = \frac{1}{N}\sum_{j=1}^Np_{j|\mathrm{past}}$. This score can be used to estimate the number of rounds needed to state with a certain confidence level that the average conditional extractability of the target state from all states is above some limit. With the introduced figures of merit, everything we said in the previous section can be translated to the scenario with a non-IID source, if the probability to pass a random round $p_j$ is exchanged with the conditional probability $p_{j|\mathrm{past}}$ and the average extractability is exchanged with average conditional extractability. As all conditional probabilities are independent, in the derivation of the final result given by Eq. \eqref{numberofcopiesverification}, in Eq. \eqref{eq:19} we use $\bar{\Xi}_{\mathrm{cond}}$ instead of $\bar{\Xi}$, and in Eq. \eqref{eq:20} we use a modified definition of $\bar{p}$.

\section{A framework for device-independent quantum state certification}\label{DIQSC}

In practice, there is a strong interest to verify the functionality of an uncharacterized source before using it in some information-processing protocol. The DI quantum state verification protocol we introduced above is not applicable in such a scenario for two reasons. Firstly, as we explained, it is not the functionality of the source that is verified, but the average extractability of the target state from the produced copies. Secondly, all the copies are consumed in the process of verification, and they cannot be reused.

In this section, we address this issue and construct a protocol for DI certification: from the copies emitted by the source, some fraction is randomly chosen to be measured, the same way they would be measured in the DI quantum state verification, while all the remaining copies are preserved to be used in some other protocol of interest. Importantly, we show that the performance of the measured copies in the quantum state verification test allows us to say something about the average extractability of the target state from the remaining copies. The task of certifying one copy based on the previous $N$ measurements has been discussed in the fully non-IID scenario in Refs.~\cite{Zhu3,Zhu4}. While the protocol works in a full adversarial scenario, it can be used to certify just a single (out of $N$) copy. In this work, we rather consider the problem of certifying a large number of copies in a DI manner. We develop an efficient scheme for the case of independent copies, while the full non-IID solutions still remain for future investigations.  As in the scenario for the DI quantum state verification, the underlying self-testing procedure can be seen as a nonlocal game, in which for every set of inputs, the collective set of outputs is either correct or wrong. Hence, every measurement round is either passed, if the set of outputs is correct, and otherwise it is failed. 

\emph{Independent copies.}  As a starting point for this section, let us state the protocol for device-independent quantum state certification, free of IID assumption.

\vskip 0.1cm
\textbf{Protocol 2:} {Device-independent quantum state certification.}
\vskip 0.1cm

$\mathbf{1)}$ Fix the lower bound  $1 - \eta_c$ on the certificate average extractability we want to certify. This implies the lower bound $1-\eta$ on the average extractability of the whole sample, and consequently to the average success probability of the whole sample $p_{\mathrm{QM}} - \varepsilon_2$. \vspace{2mm}

$\mathbf{2)}$ Fix the allowed departure from the optimal success rate $\varepsilon_1 < \varepsilon_2$, i.e., $P\geq p_1= p_{\mathrm{QM}}- \varepsilon_1$ and the corresponding verification confidence $1-\delta$. \vspace{2mm}

$\mathbf{3)}$ {Fix the number of samples $N$ according to the lower bound given in Eq. \eqref{numofcop}}.\vspace{2mm}

$\mathbf{4)}$ Run the protocol: randomly choose $N_1\approx \mu N$ instances of the given quantum system, measure them as in DI quantum state verification protocol. \vspace{2mm} 

$\mathbf{5)}$ The number of successful rounds is $q_1$. If $q_1/N_1\geq p_1$, the protocol is successful and the average {extractability} of the certificate can be found with confidence level $1-\delta$. Otherwise, the protocol is inconclusive. 
\vskip 0.1cm

Assume that the source has produced $N$ independent copies $S = \{\sigma_1,\cdots,\sigma_N\}$. For each copy we randomly choose whether it will be measured or not, for example by flipping a $\mu$-biased coin. In the end, a fraction $N_1 \approx \mu N$ of them is chosen for the verification process, and we denote this set with $I$.
The set of indices labeling the remaining $N_2=N-N_1$ copies is denoted with $J$. The $N_1$ chosen copies undergo the same procedure we describe in the DI quantum state verification. In each round, the boxes provide a correct $(p_j = 1)$ or wrong $(p_j = 0)$ answer, hence one can regard the measurement process as revealing the output of a dichotomic random variable.
The measurement phase is characterized by the number of times the boxes provided a correct answer, which we denote with $q_1$. Thus the overall verification success is  $P= \sum_{j \in I}p_j/N_1=q_1/N_1$. The certification is said to be successful if $P \geq p_{\mathrm{QM}} - \varepsilon_1 \equiv p_1$. Here $p_{\mathrm{QM}}$ denotes the quantum bound achieved for the target state. 
The aim is to show that if the average extractability $\bar{\Xi}$ is upper bounded by $1-\eta$, i.e., the success probability is upper bounded by $p_{\mathrm{QM}}-\varepsilon_2 \equiv p_2$, the verification task can be passed with exponentially small probability. Note that, in order to derive exponential tail bounds on the certification probability, we assume $\varepsilon_2 > \varepsilon_1$.
Namely, having $N$ copies at our disposal, the probability to achieve success, i.e., to have $P\geq p_1$, given average extractability $\bar{\Xi} \leq 1-\eta$, i.e., $\bar{p}\leq p_2$ [argumentation on the level of sets is the same as in quantum state verification section, see discussion around Eq. \eqref{eq:21}], is given with the following Chernoff-like bound (for the proof, see Appendix \ref{proof}):
\begin{eqnarray}\label{mmain}
&p&\left[P \geq p_1|\bar{\Xi}\leq 1-\eta\right] \leq\\ \nonumber \leq&p&\left[P \geq p_1|\bar{p}\leq p_2\right]\leq  
\left[1-\mu + \mu\mathrm{e}^{-D(p_1||p_2)}\right]^N.
\end{eqnarray}

From here, we can estimate the number of copies 

\begin{equation}\label{numofcop}
    {N\geq \frac{\ln{\delta}}{\ln{(1-\mu +\mu \mathrm{e}^{-D(p_1||p_2)})}},}
    \end{equation}
where $1-\delta$ is the confidence level. 
Now we derive lower bounds on the success probability of the remaining copies with the confidence level $1-\delta$. 
The success probability of the whole sample is

\begin{equation}
    \bar{p}= \frac{p_v N_1+p_c (N-N_1)}{N},
\end{equation}

where $p_v=\sum_{j\in I}p_{\eta_j}/N_1$ is the success probability of measured states and $p_c = \sum_{j\in J}p_{\eta_j}/(N-N_1)$ the success probability of the remaining ones. Given that trivially $p_v \leq p_{\mathrm{QM}}$, the expression $\bar{p} \geq p_2$ implies 

\begin{equation} \label{pc}
p_c \geq  {\frac{Np_2 - N_1 p_{\mathrm{QM}} }{N-N_1} = p_{\mathrm{QM}}-\frac{\epsilon_2 N}{N-N_1}\approx  p_{\mathrm{QM}}-\frac{\epsilon_2}{1-\mu} }.
\end{equation}

Due to the existence of the self-testing robust bounds (e.g. 
$p_{\eta_c}=p_{\mathrm{QM}}-c\eta_c$), we can conclude that the remaining, unmeasured copies have the average extractability {$\bar{\Xi}_c$} higher than $1-{\eta}_c$, with confidence level $1-\delta$, under the certification condition demanding that a randomly chosen sequence of measured quantum systems passed successfully $q_1$ rounds in the verification part.

Again, we discuss the case of the algebraic and nonalgebraic bound separately.
In the case of $p_{\mathrm{QM}}=1$, Taylor expansion of eq. \eqref{numofcop} gives the optimal sample efficiency $N=O(\ln{\delta^{-1}}/\mu (1-\mu) c\eta_c)$. 
On the other hand, for a nonalgebraic bound, i.e., $p_{\mathrm{QM}}<1$, one obtains sample efficiency of $N = O[\ln{\delta^{-1}}/\mu (1-\mu)^2 c ^2 {\eta}_c^2]$. Details of the expansions are presented in Appendix \ref{proof}. 

\vskip 0.2cm
\textbf{Result 2:} {DI quantum state certification.}
\vskip 0.1cm
\emph{The entangled state $\ket{\psi}$ can be device-independently certified  if there exists a robust self-test for it, based on the Bell inequality, which can be phrased as a nonlocal game. The sample efficiency is $N=O(\ln{\delta^{-1}}/c {\eta}_c)$ in the case of a Bell inequality whose quantum bound is also the algebraic bound and $N=O(\ln{\delta^{-1}}/c^2 {\eta}_c^2)$ in the case of a nonequivalence between the quantum and the algebraic bound.}
\vskip 0.1cm

\begin{figure}[h!]
    \centering
    \includegraphics[width=8.5cm]{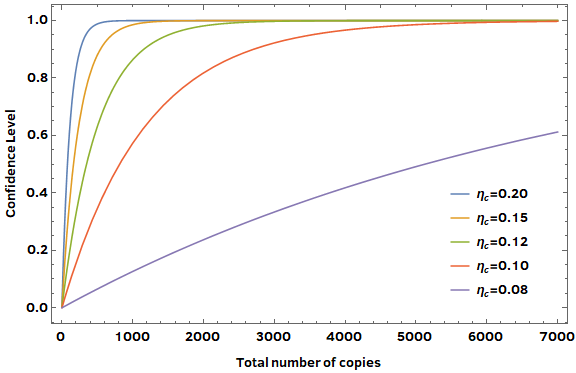}
    \caption{The average extractability of the GHZ state from the certificate in the case of a tripartite GHZ state. The verification phase is done using {a} Mermin test. Confidence\\ level for certifying average extractability of the GHZ state from the remaining copies $S_c$ is greater than $\;\;\;\;\;\;1-\eta_c$ depending on the total number of copies $N$. Here $\mu=\frac{1}{2}$ and the success rate in the verification phase is set to $0.98$.}
    \label{FidCert}
\end{figure}

We give a concrete example of certification of a tripartite GHZ state through {a} Mermin test. In our numerical simulations, we take  $\mu=1/2$ and $N_1\approx N/2$ and we set the overall verification success to $0.98$.
We provide an estimation of the average extractability $1-{\eta}_c$ of the GHZ state from the remaining copies $S_c$. Figure \ref{FidCert} illustrates how confidence level grows with a total size of the sample (number of copies) if we want to certify that average extractability of the remaining copies is greater than some value. By making this number lower, i.e., having a smaller success rate in the verification phase, the size of the sample for the reliable certification grows and the bound on the average extractability of the GHZ state from the certificate decreases for the same confidence level.

\section{Discussion}

In this paper, we developed a protocol for sample-efficient device-independent quantum state verification and certification. The procedures merge well-explored protocols for quantum state verification and self-testing. 
The biggest room for improvement is the verification procedure in fully non-IID certification scenario. In the verification scenario we tackled the problem by introducing the conditional fidelities, but it remains an open question whether some nontrivial statements can be made about the full state produced by the source. We managed to adapt the procedure to the certification scenario where not all copies are consumed through verification in cases where different copies are not identical, but are independent from each other. While trying to build a fully non-IID DI certification protocol we encountered technical difficulties whenever we aimed to certify the average extractability for more than one remaining copy. In the situation where all copies but one are measured fully non-IID DI protocol could be constructed by using the approach of conditional extractability. The aim is to build such a protocol in the near future.
For future research, we are left to explore the possibility of building a certification protocol in a strictly sequential scenario, where resorting to permutational invariance is not allowed. We hope that this might be possible with the help of entropy accumulation theorems \cite{dupuis2016entropy,Arnon_Friedman_2019}. Let us also note that our methods could be used for DI verification of resources different than states, for example, quantum gates, by merging quantum verification~\cite{Zhu_2020} and self-testing methods \cite{Sekatski_2018}. Finally, a trivial consequence of our work is sample-efficient nonlocality detection. While our work focused on the certification of quantum states, if one aims to simply detect nonlocality, it is possible to do it with a very few copies of the state, as it was similarly done for entanglement in Refs. \cite{Dimic1,Saggio_2019}. {Similar work has been done in Ref. \cite{zhang2011asymptotically}, where optimal sample efficiency was achieved for detecting nonlocality. The same method was adapted for entanglement detection \cite{zhang2013efficient} and confidence-interval construction \cite{wills2017performance}. The method presented in Ref. \cite{zhang2011asymptotically} works with early stopping rules, such that the number of samples to be used needs not to be fixed in advance. }

\section{Acknowledgements}

We acknowledge fruitful discussions with Dragoljub Go\v{c}anin and Jean-Daniel Bancal. We thank Boris Bourdoncle and Flavio Baccari for their valuable comments about the paper. We especially thank Yunguang Han for pointing out the mistake in Fig. \ref{fig2}(a) in the first version of our work. {A.G.} acknowledges the funding provided by the Faculty of Physics, University of Belgrade, through the grant by the Ministry of Education, Science, and Technological Development of the Republic of Serbia. {A.G.} acknowledges Grant No. FQXi-MGA-1806 that supported her stay in Vienna. {A.G.} would also like to thank the University of Vienna and the University of Geneva for hospitality during her stay. I.\v{S}. acknowledges support from SNSF (Starting Grant DIAQ) and funding from Starting ERC Grant QUSCO. B.D. acknowledges support from the Austrian Science Fund (FWF) through BeyondC-F7112.

\begin{appendix}
\section{{Proof of Chernoff bounds for device independent quantum state verification and certification}}\label{proof}

Let us take $N$ copies of the quantum system produced by an untrusted source. In our protocol we assume that copies are independent $\sigma_1, \; \sigma_2,...,\; \sigma_N$, but not necessarily identically distributed. We denote with $p_{\eta_j}=q-c\eta_j$ maximal probability for the $j$th copy to pass the round, where $1-\eta_j$ is extractability of the given copy and $c$ the constant that depends on the self-test. The average probability of success is given by $\bar{p}=\frac{1}{N}\sum_{j=1}^{N} p_{\eta_j}$.
In order to obtain a certificate, we measure part of the sample and compare the result with the quantum bound $p_{\mathrm{QM}}$. Quantum bound $p_{\mathrm{QM}}$ represents the probability of success of the target quantum state $|\psi\rangle$. We set $\mu$ to be the probability for a copy to be measured. On average, the experiment consumes $N_1 \approx \mu N$ copies for verification. The remaining $N_2=N-N_1$ copies are left as a certificate.
The protocol can be summarized in the following steps:
\begin{itemize}
    \item In each run we toss a $\mu$-biased coin and we get a sequence of bits $\{m_1,m_2,...,m_N\}$. If $m_j=1$ we measure the $j$th copy and obtain result $o_j$; otherwise we leave it as a (potential) certificate. The total number of measured copies is $N_1=\sum_{j=1}^N m_j$.
    \item In order to extend our analysis to all copies, we artificially define a measurement result for both measured and unmeasured copies by $\tilde{o}_j=m_j o_j$, $j=1,2,...,N$.
    \item The success rate in the verification phase can be written as $P=\frac{1}{N_1}\sum_{j=1}^{N} \tilde{o}_j$.
    \item The test is successful if $P>p_{\mathrm{QM}}-\varepsilon_1$.
\end{itemize}
Suppose that our sample has the average extractability $\bar{\Xi}$, which is upper bounded by $1-\eta$, consequently the average success probability of the whole sample $\bar{p}$ is smaller than $p_{\mathrm{QM}}-c \eta = p_{\mathrm{QM}}-\varepsilon_2$. Our target quantity is the upper bound on probability to obtain success rate of the measured part greater than $p_{\mathrm{QM}}-\varepsilon_1$, under assumption $\Xi\leq 1-\eta$, i.e., $\bar{p}\leq p_{\mathrm{QM}}-\epsilon_2$. Here $\varepsilon_1$ and $\varepsilon_2$ are free parameters lying in the interval $(0,1)$. In what follows, we see that a valid expression can only be derived for $\varepsilon_2>\varepsilon_1$. 

 For the sake of compactness, we set $p_1=p_{\mathrm{QM}}-\varepsilon_1$ and $p_2=p_{\mathrm{QM}}-\varepsilon_2$  and write certification condition as $p\left[P\geq p_1|\bar{p}\leq p_2\right]$.
 Probability to pass the test given average success probability $\bar{p}\leq p_2$ is given by
 
 \begin{align*}
     &p\left[P\geq p_1 |\bar{p}\leq p_2\right]=\\
     & = p\left[\tilde{o}_1+...+\tilde{o}_N\geq (m_1+...+m_N)p_1|\bar{p}\leq p_2\right]\\
     & = p\left[(m_1 o_1-m_1 p_1)+...+(m_N o_N - m_N p_1) \geq 0|\bar{p}\leq p_2 \right]\\
     & = p\left[s_1+...+s_N\geq 0|\bar{p}\leq p_2\right]\\
     & = p\left[\mathrm{e}^{(s_1+...+s_N)t}\geq 1|\bar{p}\leq p_2\right],
 \end{align*}
 
 for $t>0$. We {introduce} random variables $s_k=m_k o_k - m_k p_1$ with possible values $\{0, -p_1, 1-p_1\}$ and respective probabilities $\{1-\mu, \mu(1-p_k), \mu p_k\}$.
For the purpose of an example, we have $m_k=1$ with probability $\mu$, while $o_k=1$ with probability $p_k$, thus outcome $s_k=1-p_1$ appears with probability $\mu p_k$.
 Then, we apply Markov's inequality and given that $s_k$s are independent random variables (fourth line {below}):
 
 \begin{align*}
     &p\left[\mathrm{e}^{(s_1+...+s_N)t}\geq 1|\bar{p}\leq p_2\right]\leq\\ &\leq \mathbb{E}\left[\mathrm{e}^{t(s_1+...+s_N)}\right]\\ &=\Pi_{k=1}^N \mathbb{E}\left[\mathrm{e}^{t(m_k o_k-m_k p_1)}\right]\\
     &=\Pi_{k=1}^N[1-\mu +\mu(1-p_k) \mathrm{e}^{-p_1t}+\mu p_k \mathrm{e}^{(1-p_1)t}]  \\
  &\leq\left\{\frac{1}{N} \sum_{k=1}^N[1-\mu +\mu(1-p_k) \mathrm{e}^{-p_1t}+\mu p_k \mathrm{e}^{(1-p_1)t}]\right\}^N\\
&=  \left\{[1-\mu +\mu(1-\bar{p}) \mathrm{e}^{-p_1t}+\mu \bar{p} \mathrm{e}^{(1-p_1)t}]\right\}^N\\
&\leq  \left\{1-\mu+\mu \mathrm{e}^{-p_1t} + \mu p_2[\mathrm{e}^{(1-p_1)t}-\mathrm{e}^{-p_1 t}]\right\}^N.
 \end{align*}

The fifth line is the consequence of the relation between the arithmetic and the geometric means. The sixth line follows from $\sum_k p_k=N\bar{p}$, while the seventh line is a consequence of the condition $\bar{p}\leq p_2$.

The function $f(t)=1-\mu+\mu \mathrm{e}^{-p_1t} + \mu p_2 (\mathrm{e}^{(1-p_1)t}-\mathrm{e}^{-p_1 t})$ attains the minimal value for $t=\log\frac{p_1(1-p_2)}{(1-p_1)p_2}$ or equivalently $\mathrm{e}^{t}=\frac{p_1(1-p_2)}{(1-p_1)p_2}$, which after substitution gives
\begin{eqnarray}\label{maineq}
p_{\mathrm{max}}\left[P\geq p_{\mathrm{QM}}-\varepsilon_1|\bar{p}\leq p_{\mathrm{QM}}-\varepsilon_2\right] \leq \\ \nonumber 
\left[1-\mu + \mu\mathrm{e}^{-D(p_{\mathrm{QM}}-\varepsilon_1||p_{\mathrm{QM}}-\varepsilon_2)}\right]^N.
\end{eqnarray}
 Here $D(a||b) = a\log(a/b) + (1-a)\log((1-a)/(1-b))$ is the Kullback-Leibler divergence. {It is worth mentioning that the derivation of equation \eqref{maineq} follows the idea of deriving Hoeffding's inequality \cite{hoeffding1994probability}.} Finally, let us discuss the obtained result in different situations.

\textbf{Verification.} If we take $\mu=1$, which means that all copies are measured, we obtain 
 \begin{equation}
 p_{\textrm{max}}\left[P\geq p_1|\bar{p}\leq p_2\right]\leq \mathrm{e}^{-D(p_1||p_2)N}, 
 \end{equation}
which gives bound \eqref{generalverification} from the main text. Having fixed confidence level $1-\delta$, we easily estimate the average number of copies given by \eqref{numberofcopiesverification}. In order to recover scaling, we find the Taylor expansion of Eq. \eqref{numberofcopiesverification} in the limit of small $\varepsilon_1$ and $\varepsilon_2$. 
We obtain two different behaviors for algebraic ($p_{\mathrm{QM}}=1$) and nonalgebraic bound ($p_{\mathrm{QM}}<1$).
For $p_{\mathrm{QM}}=1$, we have 
\begin{equation}
N\approx \frac{\ln{\delta^{-1}}}{\varepsilon_2}\left(1+\frac{\varepsilon_1}{\varepsilon_2}\right) + o(1),
\end{equation}
where $\varepsilon_1<\varepsilon_2$. Usually, $\varepsilon_1$ and $\varepsilon_2$ are of the same order of magnitude, e.g. we can set $\varepsilon_1=\varepsilon_2/2$, giving $N=O(\ln{\delta^{-1}}/\varepsilon_2)=O(\ln{\delta^{-1}}/c \eta)$. Here we employ the relation $\varepsilon_2=c \eta$.
In the case of $p_{\mathrm{QM}}<1$, we obtain
\begin{equation}
N \approx \frac{2 (1-p_{\mathrm{QM}})p_{\mathrm{QM}} \ln{\delta^{-1}}}{\varepsilon_2^2}\left(1+\frac{2\varepsilon_1}{\varepsilon_2}\right)+ o\left(\frac{1}{\varepsilon_2}\right),
\end{equation}
which recovers $N=O(\ln{\delta^{-1}}/\varepsilon_2^2)=O(\ln{\delta^{-1}}/c^2\eta^2)$.\\
\textbf{Certification.} In any other case $\mu < 1$. Again, by fixing $\varepsilon_1<\varepsilon_2$ and taking confidence level $1-\delta$, we can estimate the number of copies, Eq. \eqref{numofcop}, necessary for certification protocol. If the protocol is successful, we have a lower bound on the average success probability $\bar{p}$ and an estimate for the lower bound on the success probability of the remaining copies $p_c$, with confidence level $1-\delta$. We have
\begin{equation}\label{apprel}
\bar{p}=\frac{N_1 p_v + (N-N_1)p_c}{N},
\end{equation}
where $p_v$ is the average success probability of the measured copies.  
If $\bar{p}\geq p_2$, directly from Eq. \eqref{apprel} and from the logical bound $p_v \leq p_{\mathrm{QM}}$ we obtain
\begin{equation}
p_c \geq {\frac{Np_2 - N_1 p_{\mathrm{QM}} }{N-N_1} = p_{\mathrm{QM}}-\frac{\epsilon_2 N}{N-N_1}},
\end{equation}
which directly translates to average certificate extractability
\begin{equation}
1-\eta_c\geq {1-\frac{N \varepsilon_2}{c(N-N_1)}} \approx 1-\frac{ \varepsilon_2}{c(1-\mu)}
\end{equation}
with confidence level greater than 
\begin{equation}
1-\delta = 1-\left[1-\mu + \mu\mathrm{e}^{-D(p_{\mathrm{QM}}-\varepsilon_1||p_{\mathrm{QM}}-\varepsilon_2)}\right]^N.
\end{equation}
For completeness, we make Taylor expansion of Eq. \eqref{numofcop} and comment on the scaling. 
In the case of $p_{\mathrm{QM}}=1$, Taylor expansion gives
\begin{equation}
N\approx \frac{\ln{\delta^{-1}}}{\mu\varepsilon_2}\left(1+\frac{\varepsilon_1}{\varepsilon_2}\right) + o(1),
\end{equation}
which after introducing $\varepsilon_2\geq c \eta_c (1-\mu)$ and $\varepsilon_1<\varepsilon_2$ recreates optimal scaling $N=O[\ln{\delta^{-1}}/\mu (1-\mu) c\eta_c]$.
Similarly, for $p_{\mathrm{QM}}<1$, after expansion we obtain
\begin{equation}
N \approx \frac{2 (1-p_{\mathrm{QM}})p_{\mathrm{QM}} \ln{\delta^{-1}}}{\mu\varepsilon_2^2}\left(1+\frac{2\varepsilon_1}{\varepsilon_2}\right)+ o\left(\frac{1}{\varepsilon_2}\right), 
\end{equation}
which gives quadratically worse scaling $N=O[\ln{\delta^{-1}}/\mu (1-\mu)^2 c ^2 {\eta}_c^2]$.

\end{appendix}

\bibliographystyle{unsrturl}
\bibliography{Biblio}

\newpage
\end{document}